\documentclass[useAMS,usenatbib]{mn2e}
\usepackage{graphicx}
\usepackage{natbib}

\def\aap{A\&A}
\def\apj{ApJ}

\def\apjs{ApJS}
\def\mnras{MNRAS}

\def\etal{et al.}
\def\nat{{\em Nature}}

\renewcommand{\vec}[1]{\mbox{\boldmath$#1$}}
\newcommand{\cn}{{\rm cn}}
\newcommand{\D}{\displaystyle}
\newcommand{\DF}[2]{\frac{\D#1}{\D#2}}

\usepackage[dvips]{color}

\begin{document}

\title[Emission from warped discs]
{Broad reprocessed Balmer emission from warped accretion discs}
\author[Wu, Wang and Dong]
{Sheng-Miao Wu\thanks{E-mail:shengmwu@mail.ustc.edu.cn}, Ting-Gui Wang
\thanks{E-mail:twang@ustc.edu.cn } and Xiao-Bo Dong\thanks{E-mail:
xbdong@ustc.edu.cn } \\
Centre for Astrophysics, University of Science and Technology of
China, Hefei, 230026, China}

\maketitle

\date{Accepted . Received ; in original form}

\markboth{Wu et al.: Emission from warped discs}{}

\begin{abstract}
Double peaked broad emission lines in active galactic nuclei are 
generally considered to be formed in an accretion disc. In this 
paper, we compute the profiles of reprocessing emission lines from 
a relativistic, warped accretion disc around a black hole in order to 
explore the possibility that certain asymmetries in the double-peaked 
emission line profile which can not be explained by a circular Keplerian 
disc may be induced by disc warping. The disc warping also provides 
a solution for the energy budget in the emission line region because it 
increases the solid angle of the outer disc portion subtended to the inner 
portion of the disc. We adopted a parametrized disc geometry and a central 
point-like source of ionizing radiation to capture the main characteristics 
of the emission line profile from such discs. We find that the ratio between 
the blue and red peaks of the line profiles becoming less than unity can 
be naturally predicted by a twisted warped disc, and a third peak can be 
produced in some cases.  We show that disc warping can reproduce the main 
features of multi-peaked line profiles of four active galactic nuclei from 
the Sloan Digital Sky Survey.
\end{abstract}


\section{Introduction}
\label{intro}

A small fraction of active galactic nuclei~(AGN) show double-peaked
broad emission line profiles \citep*{era94,era03,str03,gez07}. The
possibility has been considered for a long time that at least some
of these lines arise directly from the accretion discs assumed to 
feed the central supermassive black holes. The H$\alpha$ profile 
observed in the spectrum of Arp 102B was modelled as arising in a 
relativistic disc by \citet{che89} and \citet*{chen89} in an attempt 
to explain the asymmetric double-peaked profiles. In these stationary 
circular relativistic disc models, Doppler boosting will make the 
blue peak of the profile higher than the red one. However, 
\citet{mil90} questioned the relativistic disc explanation for 
Arp 102B, by showing that at least at some epochs the profile 
asymmetry is reversed, with the red peak higher than the blue one 
in contrary to the prediction of homogeneous, circular relativistic 
disc models. This type of profile asymmetry has been observed
in some other double-peaked sources \citep{str03,gez07}. Thus, other
scenarios for the profile asymmetry were considered later on,
including various physically plausible processes in the accretion
flow, such as spiral shocks, eccentric discs, bipolar outflows
\citep{cha93,cha94,era95,sto97,zhe90,vei91}, as well as  a
binary black hole system \citep{beg80,gas96,zha07b}. 

In addition to the line profile asymmetry, the energetic budget for 
the line emission region has not been solved so far for the 
disc model.  It is demonstrated that the local viscous heating in the 
accretion disc is not sufficient to explain the observed H$\alpha$ 
intensity \citep{chen89,era97}, thus reprocessing of ultraviolet(UV)/X-ray 
continuum from the inner accretion disc is required. However, only 
a very small fraction of radiation from inner disc is expected to 
intercept the outer part of the disc in the case of a flat geometrically 
thin disc. Two type of processes have been proposed to increase the 
fraction of light incident on to the disc emission-line region.  \citet{chen89} 
proposed that the inner part  of the accretion disc is hot and becomes 
geometrically thick due to insufficient radiative cooling, and the X-ray 
emission from such an ion-supported torus is responsible for such energy 
input.  However, double peak emitters do not always have a low accretion 
rate \citep{zha07A,bian07}.  Furthermore, a 
detailed calculation of advection dominated accretion flow (ADAF) 
models suggests that only a small portion of X-rays actually hit 
the emission-line portion of the disc \citep{cao06}. On the other hand, 
\citet{cao06} proposed that a slow moving jet can scatter a 
substantial fraction of UV/X-ray light from the inner accretion disc back 
to the outer part of the disc.  While this may be the likely case for 
radio-loud double-peaked emitters, the majority of double-peaked emitters 
are radio quiet. 

On the other hand, warped accretion discs are believed to exist in a
variety of astrophysical systems, including X-ray binaries, T Tauri
stars, and AGN. The dynamics of warping accretion discs has been
discussed by a number of authors \citep{pap83,pal95,ogi99,nel99,nel00,nay05}.
Four main mechanisms for exciting/maintaining warping in accretion
discs have been proposed, including: tidally induced warping by a
companion in a binary system \citep{teb93,teb96,lar96}, radiation driven
or self-inducing warping \citep{pri96,pri97,mal96,mal97,mal98},
magnetically driven disc warping \citep{lai99,lai03,pfl04}, and frame
dragging driven warping \citep{bap75,kum85,arm99}. As a result,
the accretion disc in some AGN should be nonplanar. Evidence
for the existence of warped discs in AGN has been found by observations
\citep{her05,cap07}. A warped disc makes it possible for the
radiation from the inner disc to reach the outer parts, enhancing
the reprocessing emission lines. Thus, this warping can solve the
long-standing energy-budget problem because the subtending angle of the
outer disc portion to the inner part has been increased by
the warp.  It can also provide the asymmetry required for the variations
of the emission lines. The effect of disc warping on iron line
profiles treated relativistically have been made by \citet{har00} and
\citet{cad03}. \citet{bac99} studied the broad-line H$\beta$ profiles
from a warped disc. Nevertheless, his treatment is non-relativistic.
A relativistic treatment of warped disc on the Balmer lines which can
be compared with observations is still not available in the literature.

In this paper, the broad Balmer emission due to reprocessing of the
central high-energy radiation by a warped accretion disc is
investigated. Our main purpose here is to describe the effect of
disc warping on the line profiles. A parametrized disc model is
adopted. In \S\ref{method} we summarize the assumptions behind our
model, and present the basic equations relevant to our problem. We
present our results and the comparison with observations in
\S\ref{result}, and conclusion and discussion in
\S\ref{sum}.

\section{Assumptions and Method of Calculation}
\label{method} 
In this paper, we will focus on how a warped disc affects the 
optical emission-line profiles, such as H$\alpha$ and H$\beta$.  
A geometrically thin disc and a central point-like source of
ionizing radiation around a black hole are assumed. 

\subsection{Geometrical considerations}
The disc can be treated as being composed of a series of concentric
circular rings of increasing radii, and laying in different planes.
The rings interact with each other via viscous stresses. Each ring is
defined by two Eulerian angles $\beta(R,t)$ and $\gamma(R,t)$~(see Fig.1)
at radius $R$, where $\beta(R,t)$ is the tilt angle of the disc
with respect to the normal to the equatorial plane, while the twist
angle $\gamma(R,t)$ describes the orientation of the line of nodes with
respect to a fixed axis in the equatorial plane.  Note that $R$ is a
purely radial coordinate and not the cylindrical one. At each radius $R$
from the center, the disc has a unit tilt vector $\vec{l}(R,t)$ which
varies continuously with radius $R$ and time $t$. Following \citet{pri96},
in an inertial frame $\rm OXYZ$ centered on the disc, the vector
$\vec{l}(R,t)$ is given by
\begin{equation}
  \vec{l}=(\cos\gamma\sin\beta,\sin\gamma\sin\beta,\cos\beta).
  \label{betagamma}
\end{equation}
We define the normalized vector towards the observer as
\begin{equation}
  \vec{i}_{obs}=(\sin i,0,\cos i),
\end{equation}
where $i$ is the angle between the line of sight and the normal to
the equatorial plane lie in the $\rm XZ$ plane.

\begin{figure}
\includegraphics[width=\hsize]{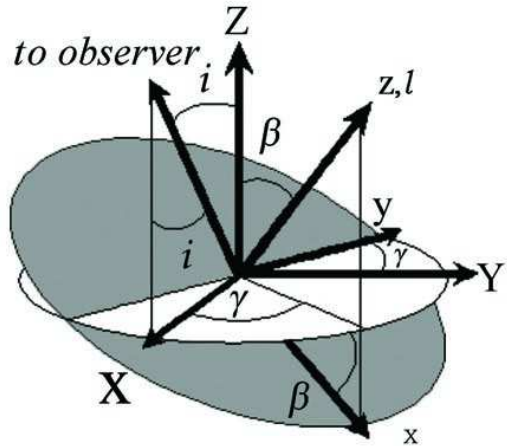}
  \caption{The geometry used in this paper. The warped accretion disc is
modeled as a collection of inclined rings, each defined by two
Eulerian angles $\beta,\gamma$ and the radial distance $R$. The
normal to the equatorial plane is aligned with the $\rm Z$
direction. The direction to the observer is denoted by $i$ that is
the angle between the direction of line of sight and the $\rm Z$-axis.
 \label{geometry}}
\end{figure}

We define the coordinates on the surface of the disc as ($R,\phi$)
with respect to a fixed Cartesian coordinate system $(x,y,z)$,
where $\phi$ is the azimuthal angle measured on the disc surface
in the direction of flow, with $\phi = \pi/2$ at the ascending
node.\footnote{This definition of $\phi$ differs by $\pi/2$ from
that of \citet{pri96}.} The position vector of a point on the ring
of radius $R$ is $\vec{R}=R\,\vec{e}_R$, where
\begin{equation}
  \vec{e}_R=\left[\matrix{\cos\gamma\cos\beta\cos\phi-\sin\gamma\sin\phi\cr
  \sin\gamma\cos\beta\cos\phi+\cos\gamma\sin\phi\cr
  -\sin\beta\cos\phi}\right]
\end{equation}
is the radial unit vector. Note that $\vec{l}\cdot\vec{e}_R=0$,
since points on the ring lie in the plane that passes through the
origin and is orthogonal to $\vec{l}$. Now $(R,\phi)$ are (in
general) non-orthogonal coordinates on the surface of the disc.
The element of surface area is
\begin{equation}
  {\rm d}\vec{S}=\left({{\partial\vec{R}}\over{\partial R}}\,{\rm d}R\right)
  \times\left({{\partial\vec{R}}\over{\partial\phi}}\,{\rm d}\phi\right),
\end{equation}
by calculating these partial derivatives directly and after some
algebra which simplifies to
\begin{equation}
  {\rm d}\vec{S}=\left[\vec{l}+(R\beta'\cos\phi+R\gamma'\sin\beta\sin\phi)\,
    \vec{e}_R\right]R\,{\rm d}R\,{\rm d}\phi,
\end{equation}
where the primes indicate differentiation with respect to $R$, and
hence, since $\vec{l}\cdot\vec{e}_R=0$, that
\begin{equation}
  |{\rm d}\vec{S}|=\left[1+(R\beta'\cos\phi+R\gamma'\sin\beta\sin\phi)^2
   \right]^{1/2}R\,{\rm d}R\,{\rm d}\phi.  
\end{equation}
The radiation flux ${\rm L}$ received from the central parts by
the unit area is proportional to
\begin{equation}
  {\rm L} \propto \DF{|(R\beta'\cos\phi+R\gamma'\sin\beta\sin\phi)|}
  {R^2\left[1+(R\beta'\cos\phi+R\gamma'\sin\beta\sin\phi)^2\right]^{1/2}},
\end{equation}
if the element in question is not obscured by other interior parts 
of the disc, otherwise the flux is set to zero~(see below). The
Balmer lines in the illuminated surface is formed as a result of
photoionization by UV/X-ray radiation from inner accretion disc. We
assume that Balmer line intensity is proportional to the radiation
being intercepted by the disc, thus the line emissivity
$\varepsilon$ on the disc surface can take the form:
$\varepsilon \propto {\rm L}$. A detailed treatment of line 
emission requires solving the vertical structure as well as the 
radiative transfer in the disc explicitly, which is beyond the scope 
of this paper. We also assume that the line emission is isotropic in the 
comoving frame. 

An alternative useful formalism being used to describe the shape of the
warped disc is introducing the scale-height of the disc which is the
displacement in the direction normal to the equatorial plane. The 
scale-height denoted by $h(r,\varphi)$ is a function of radius $r$ and
azimuthal angle $\varphi$ in cylindrical coordinates with respect to
Cartesian coordinate system ${\rm (X,Y,Z)}$. The two formalisms described
above can be related by the formula
\begin{eqnarray}
    h(r,\varphi)&=&-R\sin\beta\cos\phi\nonumber \\
       &=&-R\sin\beta\cos\psi\big/\sqrt{1-\sin^2\beta\sin^2\psi}\nonumber \\
       &=& -r\tan\beta\cos\psi.
    \label{scaleh}
\end{eqnarray}
Where $\psi = \varphi-\gamma$ and satisfies
\begin{equation}
\tan\phi\,=\,\tan\psi\cos\beta.
\end{equation}

Since it is difficult to calculate analytically the disc distortion induced 
by combined action of external torques mentioned above, here we
do not consider the disc dynamics, and merely parametrize the
disc geometry to capture the main characteristics of emission-line profiles 
from  warped discs.  In this paper, a steady disc is
adopted, and the form of the two angles $\gamma$, $\beta$ as a function of 
radial $R$  are given by
\begin{eqnarray}
  \gamma &=& \gamma_0 + n_1e^{n_2\frac{R_{in}-R}{R_{out}-R_{in}}},\\
  \beta &=& n_3\sin\left[\frac{\pi}{2}\left(\frac{R-R_{in}}{R_{out}-R_{in}}
         \right)\right].  \label{gam-beta}
\end{eqnarray}
Where $R_{in}$ and $R_{out}$ are the disc inner and outer
radii, $n_i~(i=1,2,3)$ are three free parameters used to describe the
warped disc geometry, $\gamma_0$ is a constant which is used to describe 
the longitude of the observer with respect to the coordinate system of the disc. 
This angle is mathematically equivalent to changing observational time for 
possible precession. Thus, the line profiles with different $\gamma_0$ can be 
compared with the data observed at different times.

\subsection{Photon motions in the background metric}
\label{equat} 
The propagation of radiation from the disc around a Kerr
black hole and the particle kinematics in the disc were studied by many
authors \citep*{car68,bar72}. 
We review properties of the Kerr metric and formulae for
its particle orbits, and summarize here the basic equations relevant
to this paper. Throughout the paper we use units in which $G = c = 1$, 
where $G$ is the gravitational constant, $c$ is the speed of light. 
In Boyer-Lindquist coordinates, the Kerr metric is given by
\begin{eqnarray}
ds^{2} & = & -e^{2\nu}dt^{2}+e^{2\psi}(d\phi-\omega
dt)^2+\frac{\Sigma}{\Delta}dr^{2}+\Sigma d\theta^{2} ,
\end{eqnarray}
where
\[e^{2\nu}=\Sigma\Delta/A,\,e^{2\psi}=\sin^{2}\theta A/\Sigma,\,\omega=2Mar/A,\]
\[\Sigma=r^{2}+a^{2}\cos^{2}\theta,\,\Delta=r^{2}+a^{2}-2Mr,\]
\[A=(r^{2}+a^{2})^{2}-a^{2}\Delta\sin^{2}\theta.\]
Here $M$, $a$ are the black hole mass and specific angular momentum, respectively.

The general orbits of photons in the Kerr geometry can be expressed
by a set of three constants of motion \citep{car68}. Those are the
energy at infinity $E$, the axial component of angular momentum
$E\lambda$, and carter's constant ${\cal Q}\,({=}q^2E^2)$. The
4-momentum of a geodesic has components
\begin{equation}
   p_\mu = (p_{\rm t},\,p_{\rm r},\,p_\theta,\,p_\phi) = (-E,\,\pm E\sqrt{R}/
\Delta,\,\pm E\sqrt\Theta,\,E\lambda),
\end{equation}
with
\begin{eqnarray*}
    R &=& r^4 + \left(a^2-\lambda^2-q^2\right)r^2 +2M\left[q^2+(\lambda-a)^2
        \right]r - a^2 q^2 \;,  \\
    \Theta &=& q^2 + a^2 \cos^2\theta - \lambda^2 \cot^2\theta\;.
\end{eqnarray*}
From this, the equations of motion governing the orbital trajectory
can be obtained. The motion in the $r$-$\theta$ plane is governed by \citep{bar72}
\begin{eqnarray}
    \int_{r_e}^r \frac{dr}{\sqrt{R(r)}} = \pm
        \int_{\theta_e}^\theta \frac{d\theta}{\sqrt{\Theta(
       \theta)}} \;,
    \label{r-thet}
\end{eqnarray}
where $r_e$ and $\theta_e$ are the starting values of $r$ and
$\theta$. The $\phi$ coordinate along the trajectory is calculated by 
\citep{wil72,vie93}
\begin{equation}
\int_{\phi_e}^{\phi}{\rm d}\phi =
\int_{\theta_e}^\theta\frac{\lambda{\rm d}\theta}{\sin^2\theta
  \sqrt{\Theta(\theta)}} + \int_{r_e}^r \frac{(2ar-\lambda a^2){\rm d}r}
  {\Delta\sqrt{R(r)}}. 
\label{phi-thet}
\end{equation}

Consider the orbit equation in the form
\begin{eqnarray}
    \int_{r_{ms}}^r \frac{dr}{\sqrt{R(r)}} = \pm
        \int_{\pi/2}^\theta \frac{d\theta}{\sqrt{\Theta(
       \theta)}} = P \;,
    \label{e-th}
\end{eqnarray}
where $r_{ms}$ is the radius of the marginally stable orbit for a maximal 
Kerr black hole, $a=0.998$. Then $\theta$ can be expressed in terms 
of $P$ as follows \citep*{cad98}
\begin{eqnarray}
    \cos\theta(r) = \mu_+\cn(a\sqrt{\mu_+^2 + \mu_-^2}P\pm K(m)|m) \;,
    \label{thet}
\end{eqnarray}
where $m=\frac{\mu_+^2}{\mu_+^2 + \mu_-^2}$, $\mu_\pm^2$ are defined by
\begin{eqnarray}
    \mu_\pm^2 = \frac{1}{2a^2}\left\{\left[\left(\lambda^2+q^2-a^2\right)^2
        +4a^2q^2\right]^{1/2}\mp\left(\lambda^2+q^2-a^2\right)\right\} \;.
    \label{mupm}
\end{eqnarray}
and 
\begin{eqnarray}
  \hspace{-0cm}  K(m) = \cn^{-1} \left(\left.0\right|m\right).
\end{eqnarray}
Where $K(m)$ is the complete elliptic integral of the first kind.

For photons emitted from radii~($>$100$r_g$) propagating to infinity the differences 
between Kerr and Schwarzschild black holes are indistinguishable. Neglecting terms 
of order $1/r^2$ and higher, the Kerr metric reduces to the Schwarzschild metric. 
The formulae for the particle orbits in this case are simplified. 
Here we summarize the basic equations relevant to this paper. The line element in 
Schwarzschild geometry is written as follows:
\begin{equation}
{\rm d}s^{2}=-(1-2/r){\rm d}t^{2}+\frac{1}{1-2/r}{\rm d}r^{2}+
r^{2}{\rm d\vartheta}^{2}+ r^{2}{\rm sin}^{2}\vartheta{\rm
d}\phi^{2}.
\end{equation}

The observer is assumed to be located at ($r_{\rm o}, \vartheta_{\rm o},
\phi_{\rm o}$). For a approximately spherical background, each inclined ring 
of the disc can be considered lying in the equatorial plane. Owing to a 
spherically symmetric metric there is no favored direction for the black hole;
in this sense, $\vartheta_{\rm o}$ can be acted as the angle between the 
observer and the normal to the disc and is determined by taking the 
dot-products of the unit vector $\vec{l}$ of the disc with the normalized 
vector $\vec{i}_{\rm obs}$ to the observer. By definition, we get
\begin{equation}
\cos\vartheta_{\rm o} = \cos\gamma\sin\beta\sin i+ \cos\beta\cos i
\end{equation}
The angle $\phi_{\rm 0}$ can be obtained by setting the mixed
triple product of three vectors $\vec{i}_{\rm obs}, \vec{l}$ and
$\vec{e}_{\rm R}$ equal zero, $(\vec{i}_{\rm obs}\; \vec{l}\;
\vec{e}_{\rm R})=0$. From this equation, the angle is determined by
\begin{equation}
\tan\phi_{\rm o} = \frac{\sin\gamma\sin i}{\sin\beta\cos i -
\cos\beta\cos\gamma\sin i}.
\end{equation}

The photon motion equations are given by \citep{lu01}:
\begin{equation}
    \int_{r_e}^r \frac{dr}{\sqrt{R(r)}} = \pm
        \int_{\vartheta_e}^\vartheta \frac{d\vartheta}{\sqrt{\Theta(
       \vartheta)}} \;,
    \label{r-th}
\end{equation}
\begin{equation}
\int_{\phi_e}^{\phi}{\rm d}\phi =
\pm\int_{\vartheta_e}^\vartheta\frac{\lambda{\rm d}\vartheta}{{\rm
sin}^{2}\vartheta \sqrt{\Theta(\vartheta)}}. \label{phi-th}
\end{equation}
where $r_e$, $\vartheta_e$ and $\phi_e$ are the starting values
of $r$, $\vartheta$ and $\phi$.

Define $\mu = \cos\vartheta$, then in equations~(\ref{r-th}) and 
(\ref{phi-th}) the integral over $\mu$ can be worked out in terms
of a trigonometric integral
\begin{eqnarray}
  \int_{\pi/2}^\vartheta \frac{d\vartheta}{\sqrt{\Theta(\vartheta)}} &=&
  \int_0^{\mu}\frac{d\mu}{\sqrt{q^2-\left(\lambda^2+q^2\right)\mu^2}}
       \nonumber \\
       &=&\frac{1}{\sqrt{\lambda^2 + q^2}}\, \rm {sin^{-1}
       \left(\mu/\mu_+\right)},
    \label{mu_int}
\end{eqnarray}
\begin{eqnarray}
  \int_{\pi/2}^{\vartheta}\frac{\lambda{\rm d}\vartheta}{{\rm
      sin}^{2}\vartheta \sqrt{\Theta(\vartheta)}} &=&
  \int_0^{\mu}\frac{\lambda d\mu}{(1-\mu^2)\sqrt{\Theta_\mu(\mu)}}\nonumber \\
 &=& \pm \rm{sin^{-1}\sqrt{\frac{(1-\mu_+^{2})\mu^{2}}{\mu_+^{2}(1-\mu^{2})}}}.
    \label{phi_int}
\end{eqnarray}
Where\ $0\leq \mu < \mu_+$\ , \ $\mu_+ = \sqrt{q^2/(\lambda^2 +
q^2)}$\ and $\Theta_\mu(\mu) = q^2-(\lambda^2+q^2)\mu^2$. The
integral over $r$ can be worked out with inverse Jacobian elliptic
integrals \citep[see e.g.][]{cad98,wu07}.

Although our calculation is limited to Schwarzschild metric for photons 
travelling to infinity, for the radii~($>$100 $r_{\rm g}$) taken into account in 
this paper, the results are essentially the same for the Kerr black hole. 
On the other hand, for the light from inner part of the disc, a maximal Kerr metric 
is still used. 
 
\subsection{The observed line flux}

Due to relativistic effects, the photon frequency will shift from
the emitted frequency $\nu_{\rm e}$ to the observed one $\nu_{\rm o}$
received by a rest observer with the hole at infinity. We introduce
a $g$ factor to describe the shift which is the ratio of observed
frequency to emitted one:
\begin{eqnarray}
    g & = & \nu_{\rm o} / \nu_{\rm e} = (\vec{p}\cdot\vec{u}_{\rm o})/
    (\vec{p}\cdot\vec{u}_{\rm e}) \nonumber \\
      & = & (1 - \Omega\lambda)^{-1}\sqrt{1-\frac{3}{r}},
     \label{gvalue}
\end{eqnarray}
where $\Omega = r^{-3/2}$ is the Keplerian angular velocity,
$\vec{p}, \vec{u}_{\rm o}, \vec{u}_{\rm e}$ are the 4-momentum of
the photon, the 4-velocity of the observer and the emitter,
respectively.

The specific flux density $F_{\rm o}(\nu_{\rm o})$ at frequency
$\nu_{\rm o}$ as observed by an observer at infinity is defined as
the sum of the observed specific intensities $I_{\rm o}(\nu_{\rm
o})$ from all parts of the accretion disc surface, which is given
by \citep{cun75}
\begin{eqnarray}
 F_{\rm o}(\nu_{\rm o})&=&\int I_{\rm o}(\nu_{\rm o})d\Omega_{\rm obs}\nonumber\\
  & = & \int g^3 I_{\rm e}(\nu_{\rm e}) d\Omega_{\rm obs} \;.
    \label{feo1}
\end{eqnarray}
where $d\Omega_{\rm obs}$ is the element of the solid angle
subtended by the image of the disc on the observer's sky and we
have made use of the relativistic invariance of $I_{\nu}/\nu^{3}$,
$\nu$ is the photon frequency measured by any local observer on
the path.

$I_{\rm e}(\nu_{\rm e})$ is the specific intensity measured by an
observer corotating with the disc, and can be approximated by a 
$\delta$-function, $I_{\rm e}^{\prime}(\nu_{\rm e}^{\prime}) = 
\varepsilon\delta(\nu_{\rm e}^{\prime}-\nu_{\rm e})$, where 
$\varepsilon$ is the emissivity per unit surface area. From 
well-known transformation properties of $\delta$-functions we have 
$\delta(\nu_{\rm e}^{\prime}-\nu_{\rm e})=g\delta(\nu_{\rm o}-g\nu_{\rm e})$, 
using this in equation~(\ref{feo1}), we obtain
\begin{eqnarray}
    F_{\rm o}(\nu_{\rm o}) = \int \varepsilon g^4 \delta(\nu_{\rm o} - 
g\nu_{\rm e}) d\Omega_{\rm obs}\;.
    \label{feo2}
\end{eqnarray}

In order to calculate the integration over $d\Omega_{\rm obs}$,
using two impact parameters $\alpha$ and $\beta$, measured
relative to the direction to the centre of the black hole, firstly
introduced by \citet{cun73} is convenient. They are related to two
constants of motion $\lambda$ and $q$ by equations
\begin{eqnarray}
    \alpha = - \lambda/\sin\vartheta_{\rm o},\,
    \beta = \pm\left(q^2 - \lambda^2\cot^2
        \vartheta_{\rm o}\right)^{1/2} \!, \label{alp_beta}
\end{eqnarray}
The element of solid angle seen by the observer is then
\begin{eqnarray}
    d\Omega_{\rm obs} = \frac{d\alpha d\beta}{r_{\rm o}^2}
     = \frac{q}{r_{\rm o}^2\beta\sin\vartheta_{\rm o}}
     \frac{\partial(\lambda,q)}{\partial(r,g)}\;dr\;dg,
\label{solid_obs}
\end{eqnarray}
where $r_{\rm o}$ is the distance from the observer to the black
hole.

Substituting equation~(\ref{solid_obs}) into equation~(\ref{feo2})
gives the desired result:
\begin{eqnarray}
    F_{\rm o}(\nu_{\rm o}) & = & \frac{q}{r_{\rm o}^2\beta\sin\vartheta_{\rm o}}
     \int \varepsilon g^4 \delta(\nu_{\rm o}-g\nu_{\rm e}) 
     \frac{\partial(\lambda,q)}{\partial(r,g)}\;dr\;dg.
    \label{feo3}
\end{eqnarray}

In the calculation of the total flux of reprocessing emission line, shadowing
of the elements by the inner parts of the disc must be taken into account.
This has been done as follows. The contribution of each element denoted by
($r,\varphi$) to the total
line flux from the disc is calculated by equation~(\ref{feo3}) only in case
that there does not exist another element with $\cos\theta_i > 
\cos\theta(r_i,\varphi_i)$ tracing the trajectory of photon, and is zero in 
the opposite case. Here $\cos\theta_i = h(r_i<r,\varphi_i)/r_i$,  $\varphi_i$ 
and $\cos\theta(r_i,\varphi_i)$ are calculated by equations~(\ref{phi-thet}) 
and (\ref{thet}), respectively: 
\begin{eqnarray}
    \varphi_i = \varphi - \int_{r_i}^r \frac{2ar{\rm d}r}{\Delta\sqrt{R(r)}},
    \label{varphi}
\end{eqnarray}
\begin{eqnarray}
    \cos\theta(r_i,\varphi_i) = \mu_+\cn(a\sqrt{\mu_+^2 + \mu_-^2}P -  K(m)|m) \;,
\end{eqnarray}
Equation~(\ref{varphi}) is obtained by setting $\lambda=0$ in 
equation~(\ref{phi-thet}) as a approximation for a pointlike source 
assumption. In the calculation, the disc is divided into 1000 rings 
logrithmically spaced in radial direction , and we checked every increment 
of $r$ for the inequality along the ray path. In the calculation we neglect 
the contribution from the lower surface of the disc or the higher order images 
of the emitter due to the fact that at low observer inclination angles the 
extra flux is small compared to the direct images. And a more correct calculation 
will need to handle these effects correctly and may give up to a factor of two 
change in central region \citep[see e.g.][]{vie93}.

\subsection{Method of calculation }
\label{meth}

With all of the preparation described in the previous section, we
now turn to how to calculate the line profiles numerically. We
divide the disc into a number of arbitrarily narrow rings, and
emission from each ring is calculated by considering its
axisymmetry. We shall denote by $r_{\rm i}$ the radius of each
such emitting ring. For each ring there is a family of null
geodesics along which radiation flows to a distant observer at
polar angle $\vartheta_{\rm o}$ from the disc's axis. As far as a
specific emission line is concerned, for a given observed
frequency $\nu_{\rm o}$ the null geodesics in this family can be
picked out if they exist. So, the weighted contribution of this
ring to the line flux can be determined. The total observed flux
can be obtained by summing over all emitting rings.

The main numerical procedures for computing the line profiles are
as follows:
\begin{enumerate}
\item Specify the relevant disc system parameters: $r_{\rm in}, r_{\rm
out}, n_1, n_2, n_3$ and $i$, $\gamma_0$ .

\item The flux from the disc surface is integrated using Gauss-Legendre 
integration with an algorithm due to Rybicki G. B. \citep{pre92}. The routine 
provides abscissas $r_{\rm i}$ and weights $\omega_{\rm i}$ for the 
integration.

\item For a given couple~($r_{\rm i},g$) of a ring, the two constants of
motion $\lambda$ and $q$ are determined if they exist. This is done in the
following way. The value of $\lambda$ is obtained by an alternate form of
equation~(\ref{gvalue})
\begin{eqnarray}
    \lambda & = & \frac{1}{\Omega}\left(1-\frac{1}{g}\sqrt{1-\frac{3}{r}}\right),
    \label{lambda}
\end{eqnarray}
the value of $q$ is determined for solving photon trajectory
equation~(\ref{r-th}). Then the contribution of this ring on the
flux for given frequency $\nu_{\rm o}$ with respect to $g$ is
estimated. Varying $g$, this step is repeated, thus the flux
contribution of this ring to all the possible frequency is
obtained.

\item For each g, the integration over $r$ of equation~(\ref{feo3})
can be replaced by a sum over all the emitting rings weighted by $\omega_{\rm i}$
\begin{eqnarray}
    F_{\rm o}(\nu_{\rm o}) &=& \sum_{i=1}^n \frac{q \varepsilon \nu_{\rm o}^4}
    {r_{\rm o}^2 \nu_{\rm e}^4 \beta\sin\vartheta_{\rm o}}
 \left.\frac{\partial(\lambda,q)}{\partial(r,g)}\right|_{\rm r=r_i}\omega_{\rm i}.
    \label{feo4}
\end{eqnarray}
The Jacobian $[\partial(\lambda,q)/\partial(r,g)]$ in the above
formula was evaluated by a finite difference scheme. 
In the calculation the self-obscuration along the line of sight are also 
taken into account. This has been done as follows. The contribution of each element 
denoted by ($r_{\rm i},g$) to the total line flux from the disc is calculated by 
equation~(\ref{feo4}) only in the case that there does not exist another element with 
$\cos\theta_i > \cos\theta(r_i,\varphi_i)$ along the ray path, and is zero in the 
opposite case. Here $\cos\theta_i = h(r_i,\varphi_i)/r_i$, 
$\cos\theta(r_i,\varphi_i)$ calculated by:
\begin{eqnarray}
    \cos\theta(r_i,\varphi_i) &=&\left\{\begin{array}{ll} \mu_+\sin(\pi - \rm{KM} - 
   \sqrt{\lambda^2 + q^2}P),  \\
   \mu_+\sin(\rm {KM} \pm \sqrt{\lambda^2 + q^2}P),
   \end{array}\right.
\end{eqnarray}
where the first equation describes the case with one turning point in $\theta$ component 
along the geodesic, $\varphi_i$ can be calculated by equation~(\ref{phi-th}) and 
equation~(\ref{phi_int}), and $\rm {P, KM}$, $\mu_+$ satisfy 
\begin{eqnarray}
    \int_{r_i}^\infty \frac{dr}{\sqrt{R(r)}}&=& \pm
        \int_{\theta}^{i} \frac{d\theta}{\sqrt{\Theta(
       \theta)}} =\rm{P} \;,\nonumber\\[2mm]
    \rm{KM} &=& \sin^{-1}(\frac{\cos i}{\mu_+}) \;,\nonumber\\[2mm]
    \mu_+ &=& \sqrt{q^2/(\lambda^2+q^2)} \;.\nonumber
\end{eqnarray}
\end{enumerate}

From the above formula, one determines the line flux at an arbitrary frequency
$\nu_{\rm o}$ from the disc. The observed line profile as a function of frequency 
$\nu_{\rm o}$ is finally obtained in this way.

\section{Results}
\label{result}

In the accretion disc model, the double-peaked emission lines are
radiated from the disc region between around several hundred
gravitational radii to more than 2000$r_g$; here $r_{\rm g}$ is the
gravitational radius and the widths of double-peaked lines range
from several thousand to nearly 40,000~$\rm km\,s^{-1}$
\citep{wang05}. In our model, all of the parameters of the warped
disc are set to be free. The reasonable range are: disc radius from 100$r_g$ 
to 2000$r_g$, $n_1$ from 0 to $3\pi$, $n_2$ from 2 to 4, $n_2$ from 0 to 
$30^{\circ}$. This model has the same number of 
free-parameters as an eccentric disc model \citep{era95}. In the range of 
frequency from 4.32 to 4.78 in units of $10^{14}Hz$, 180 bins are used.
Considering the brodening due to electron scattering or turbulence, all our
results are smoothed by convolution with 3$\sigma$ Gaussian~(corresponding
to 500~$\rm km\,s^{-1}$).

\begin{figure*}
\includegraphics[width=0.8\hsize]{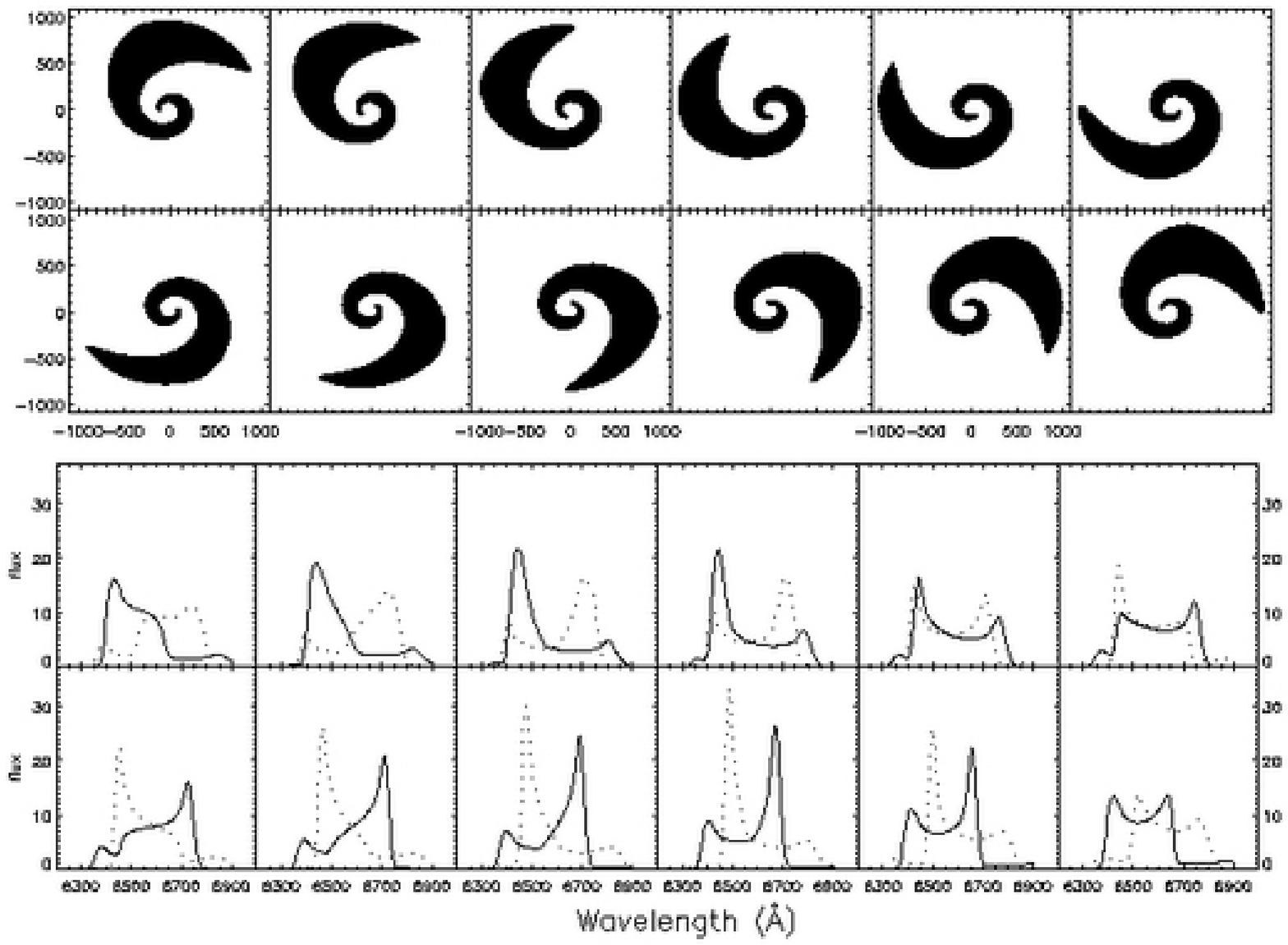}
  \caption{The images~(upper panel) of the illuminated area of the disc
and H$\alpha$ line profiles~(lower panel) computed by our code for twist
warped disc case for $i=30^{\circ},\; n_1=2.1\pi,\;n_2=3.0,\; n_3=10^{\circ}$, 
the disc zone is from $R_{\rm in}=100r_{\rm g}$ to $R_{\rm out}= 
1000 r_{\rm g}$. The profiles from both prograde disc~(black line) and 
retrograde disc~(dotted line) are shown. The longitude of the observer 
$\gamma_0$ with respect to the coordinate system of the disc takes steps of 
$30^{\circ}$ from $0^{\circ}$ to $330^{\circ}$~(from top left-hand side to 
bottom right-hand side). All the results are smoothed by convolution with 
3$\sigma$ Gaussian~(corresponding to 500~$\rm km\,s^{-1}$).
 \label{twist1}}
%\end{figure*}

%\begin{figure*}[htp]
\includegraphics[width=0.8\hsize]{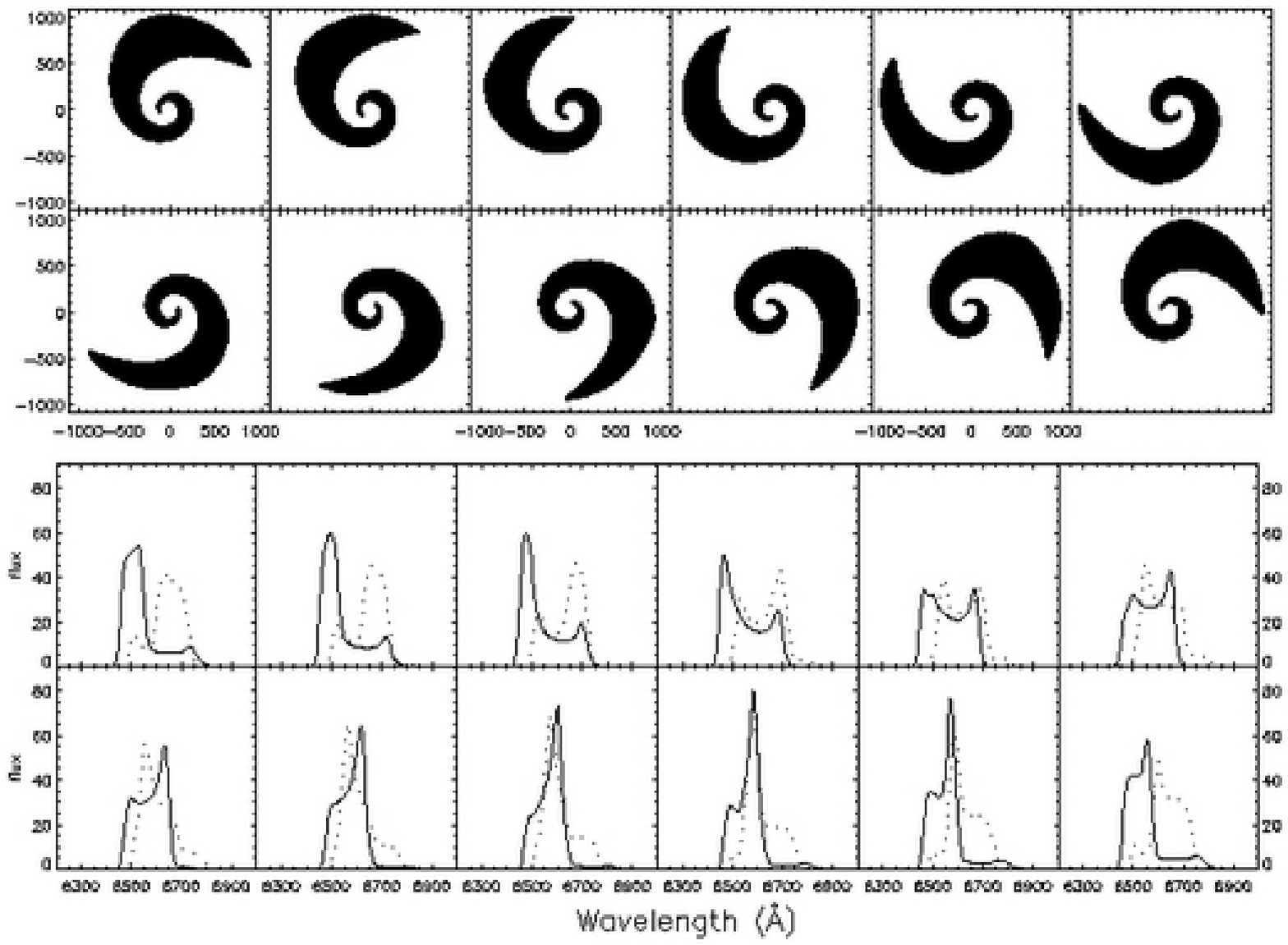}
  \caption{As in Fig.\ref{twist1} but with $i=15^{\circ},\; n_3=20^{\circ}$.
 \label{twist2}}
\end{figure*}

\begin{figure*}
\includegraphics[width=0.8\hsize]{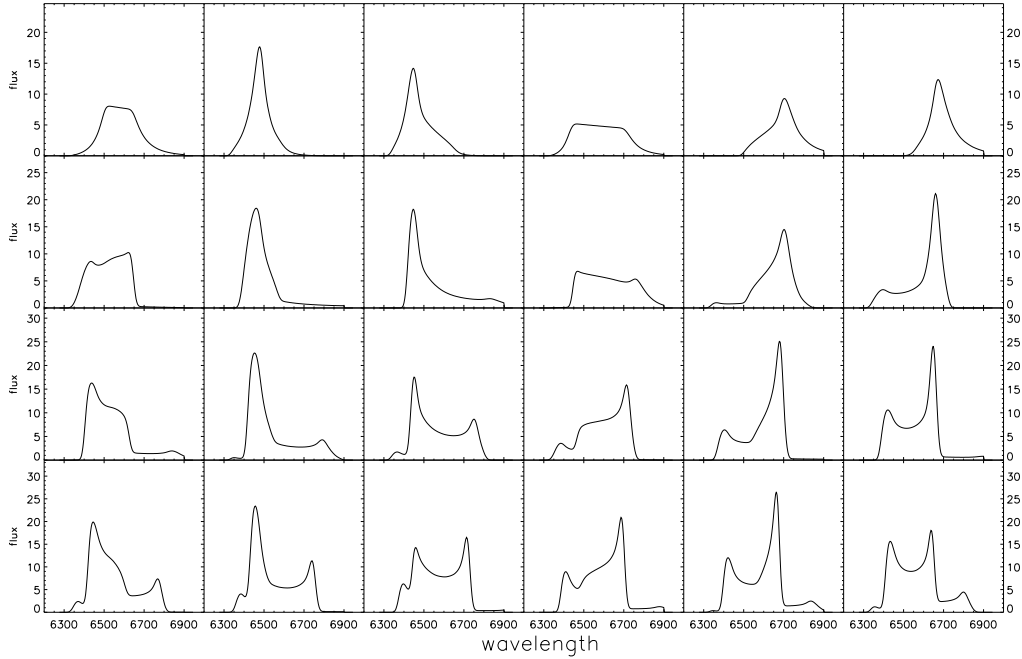}
  \caption{Comparison of the H$\alpha$ line profiles generated by our
code for warped disc cases with different phase amplitude~(described by 
parameter $n_1$), $n_1 =0,\;\pi,\;2\pi,\;3\pi$~(from top to bottom 
panels). The other parameters are: $i=30^{\circ},\; n_2=3.0,\; n_3 = 
10^{\circ},\; R_{\rm in}=100r_{\rm g}$ and $R_{\rm out}= 1200 r_{\rm g}$. 
The longitude of the observer $\gamma_0$ with respect to the coordinate 
system of the disc varies from $0^{\circ}$ to $300^{\circ}$ with a step 
of $60^{\circ}$~(from left- to right-hand side).
 \label{twist3}}
\end{figure*}

\subsection{Line profiles from twisted warping discs}
We extend the numerical code developed by \citet{wu07} to deal with
the warped discs. The images of the illuminated area and the
H$\alpha$ line profiles computed by our code for a steady-state
twisted warping disc are shown in Figs~\ref{twist1} and \ref{twist2}.
The disc zone is from $R_{\rm in}=100r_{\rm g}$ to $R_{\rm out}=
1000 r_{\rm g}$, the other parameters are $i=30^{\circ},\;
n_1=2.1\pi,\;n_2=3.0,\; n_3=10^{\circ}$, and $i=15^{\circ},\;
n_1=2.1\pi,\;n_2=3.0,\; n_3=20^{\circ}$, respectively. The longitude
of the observer $\gamma_0$ with respect to the coordinate system of
the disc from top left-hand side to bottom right-hand side varies from 
$0^{\circ}$ to $330^{\circ}$ with a step of $30^{\circ}$. The image of the 
illuminated portion of the disc has a shape similar to a one-armed disc. 
The image area of the disc being projected on the observer's sky  
from the far side of the disc due to a little angle 
between the tilt vector $\vec{l}(R,t)$ and the line of sight significantly 
larger than the near parts are indicated in Figs~\ref{twist1} and \ref{twist2}.  
In our code, one can specify an arbitrary grid for $r$ or $\phi$. In our 
calculations, all the images contain $400\times 360$ pixels in this paper. 
The line profiles drawn in black are restricted to prograde disc models~
(relative to the normal to equatorial plane or black hole spin), the 
effects of a retrograde disc~(identical to the prograde case with a reversed 
spin) on the line profiles should also be taken into account if the 
orientations of the AGN are isotropically distributed in space, those are 
drawn with dotted line in the plots.  The profiles from a prograde 
disc and its corresponding retrograde disc is close to but not exactly 
reflective symmetric about $v=0$ because of relativistic boosting, the 
gravitational redshift and the asymmetry of the disc. Note also that there  
are triple-peaked profiles similar to those found by observations~\citep{vei91} 
and the obvious variation in the red/blue peak positions.  

The influence of the phase amplitude~(described by parameter $n_1$)
on the line profiles for $n_1 =0$~(twist-free), $\pi,\;2\pi,\;3\pi$ are 
shown in Fig.~\ref{twist3}. The disc zone is from $R_{\rm in}=100r_{\rm g}$ 
to $R_{\rm out}=1200 r_{\rm g}$, the other parameters are $i=30^{\circ},\;
n_2=3.0,\; n_3=10^{\circ}$.  The longitude of the 
observer $\gamma_0$ with respect to the coordinate system of the disc 
varies from $0^{\circ}$ to $300^{\circ}$ with a step of $60^{\circ}$~
(from left- to right-hand side). For twist-free and low phase amplitude discs, 
the asymmetric as well as frequency-shifted single-peaked line profiles 
expected in most cases are produced.  For $n_1>2\pi$, there is a fairly 
large possibility for a triple-peaked line profile. The influence of the 
phase amplitude on the line profiles is remarkable. 

\begin{figure}
\includegraphics[width=1.10\hsize]{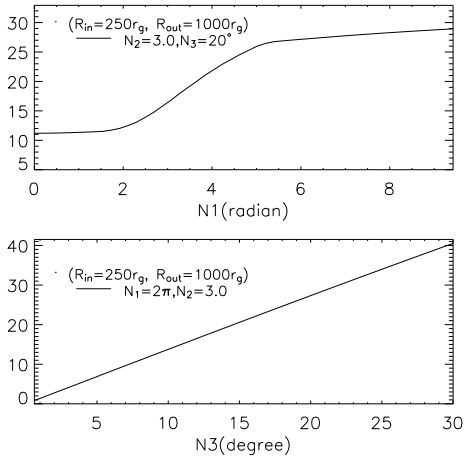}
  \caption{The fraction of light incident to the disc from the inner region 
as a function of $n_1$ or $n_3$. The disc zone is from $R_{\rm in}=250r_{\rm g}$
to $R_{\rm out}= 1000 r_{\rm g}$.
 \label{sang}}
\end{figure}

The fraction of light incident to the disc from the inner region is
a function of $n_1,n_3$. We calculate the fraction received by a
warped disc, the results are shown in Fig.~\ref{sang}. The fraction reaches 
$27.3$ per cent for a twisted warp disc $n_1=2.0\pi$, $n_2=3.0$, $n_3=20^{\circ}$.

\subsection{Comparison with the observations}

As the first step, our main purpose in this paper is to demonstrate the 
effect of disc warping on the line profiles, while the detailed fitting of 
the observational data will be presented in a future paper. From the sample 
of AGN with double-peaked Balmer lines drawn from the Sloan Digital Sky 
Survey~(SDSS) spectroscopic data set (Shan et al., in preparation), we pick 
up four sources with peculiar profiles: SDSS J084205.57+075925.5, 
SDSS J232721.96+152437.3, SDSS J094321.97+042412.0, 
SDSS J093653.84+533126.8. Two of them have explicit triple-peaked profiles, 
the others show a net shift of the emission lines to the red. These features 
are hard to interprete via a homogeneous, circular relativistic disc 
model. The SDSS spectrum of one object (SDSS J232721.96+152437.3) is 
dominated by starlight from the host galaxy; we subtract the starlight and 
the nuclear continuum with the method as described in detail in Zhou 
et al.~(2006); The other three spectra have little starlight contamination, 
the nuclear continuum and the Fe\,II emission multiplets are modelled as 
described in detail in Dong et al.~(2008). The FeII emission lines are 
apparent in J0936+5331.

\begin{figure}
\includegraphics[width=\hsize]{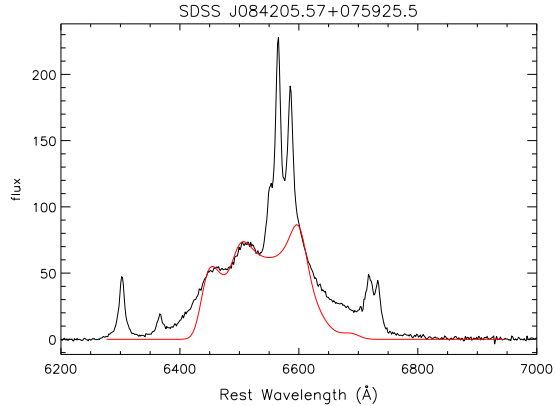}
  \caption{comparison of the H$\alpha$ line profiles computed by our 
code~(red line) with the observation of SDSS J084205.57+075925.5. The parameters 
are: $i=13^{\circ},\; n_1=3.0\pi,\;n_2=3.0,\;n_3=28^{\circ},\; \gamma_0 = 
95^{\circ}$, the disc zone is from $R_{\rm in}=250r_{\rm g}$
to $R_{\rm out}= 1000 r_{\rm g}$ for a prograde disc.
 \label{f327}}
\end{figure}

\begin{figure}
\includegraphics[width=\hsize]{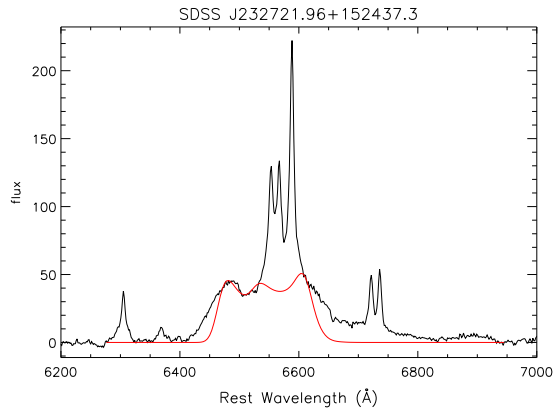}
  \caption{comparison of the H$\alpha$ line profiles computed by our 
code~(red line) with the observation of SDSS J232721.96+152437.3. The parameters 
are: $i=10^{\circ},\; n_1=2.6\pi,\;n_2=3.0,\;n_3=22^{\circ},\; \gamma_0 = 
88^{\circ}$, the disc zone is from $R_{\rm in}=250r_{\rm g}$
to $R_{\rm out}= 1000 r_{\rm g}$ for a prograde disc.
 \label{f409}}
\end{figure}

The observed broad line spectrum of sources SDSS J084205.57+075925.5,\\ 
SDSS J232721.96+152437.3 both have a third peak. The line profiles~(red 
line) calculated by our code compared with observations  are shown in 
Figs~\ref{f327} and \ref{f409}. The disc zone is from $R_{\rm in}=250r_{\rm g}$ 
to $R_{\rm out}= 1000 r_{\rm g}$, and $n_2=3.0$ for a prograde disc case. 
The other parameters are: $i=13^{\circ},\; n_1=3.0\pi, \;n_3 = 
28^{\circ},\; \gamma_0 = 95^{\circ}$ for SDSS J084205.57+075925.5, 
$i = 10^{\circ},\; n_1=2.6\pi,\;n_3 = 22^{\circ},\; \gamma_0 = 88^{\circ}$ 
for SDSS J232721.96+152437.3, respectively. We have taken the approach 
that the disk-like component should be fitted by eye to as much of the peaks 
match as possible.  While our model does not 
reproduce the extended wing,  the peak positions and heights are reasonably 
in agreement with the observed profiles.  

\begin{figure}
\includegraphics[width=\hsize]{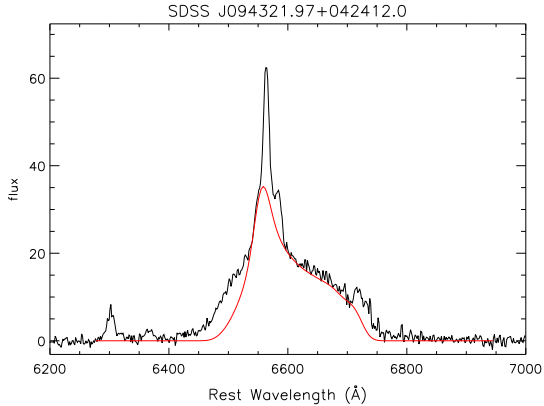}
  \caption{comparison of the H$\alpha$ line profiles computed by our 
code~(red line) with the observation of SDSS J094321.97+042412.0. The parameters 
are: $i=17^{\circ},\; n_1=2.1\pi,\;n_2=3.0,\;n_3=24^{\circ},\; \gamma_0 = 
202^{\circ}$, the disc zone is from $R_{\rm in}=250r_{\rm g}$
to $R_{\rm out}= 1000 r_{\rm g}$ for a retrograde disc.
 \label{f454}}
\end{figure}

\begin{figure}
\includegraphics[width=\hsize]{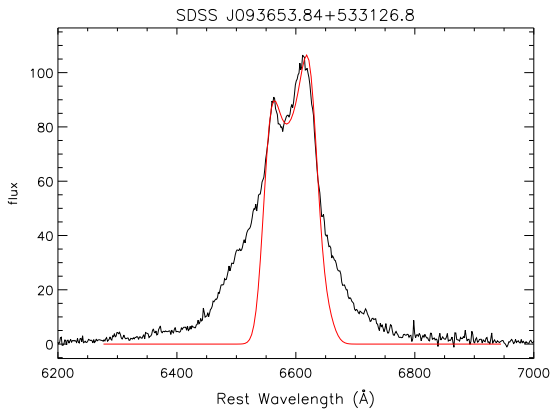}
  \caption{comparison of the H$\alpha$ line profiles computed by our 
code~(red line) with the observation of SDSS J093653.84+533126.8. The parameters 
are: $i=8^{\circ},\; n_1=2.1\pi,\;n_2=3.0,\;n_3=6^{\circ},\; \gamma_0 = 
117^{\circ}$, the disc zone is from $R_{\rm in}=250r_{\rm g}$
to $R_{\rm out}= 1000 r_{\rm g}$ for a retrograde disc.
 \label{f473}}
\end{figure}

A retrograde disc is needed to match the line profiles with the observations 
for SDSS J094321.97+042412.0, J093653.84+533126.8. Fig.~\ref{f454} shows the 
comparison between the line profiles and the observed profile for SDSS 
J094321.97+042412.0 with $i=17^{\circ},\; n_1 = 2.1\pi,\;n_3=24^{\circ},\; 
\gamma_0 = 202^{\circ}$.  And the parameters in Fig.~\ref{f473} for SDSS
J093653.84+533126.8 are: $i=8^{\circ},\; n_1=2.1\pi, \;n_3=6^{\circ},\; \gamma_0 
= 117^{\circ}$. The disc zone is still taken from $R_{\rm in}=250r_{\rm g}$  
to $R_{\rm out}= 1000 r_{\rm g}$ and $n_2=3.0$. We see that the overall match 
except that in the two wings between the line profiles and the observations is 
good qualitatively. The deviation in two wings may, mostly due to the central 
point-like source adopted, diminish the contribution from the relatively 
inner part of the reprocessing region. 

\begin{figure*}
\includegraphics[width=0.75\hsize]{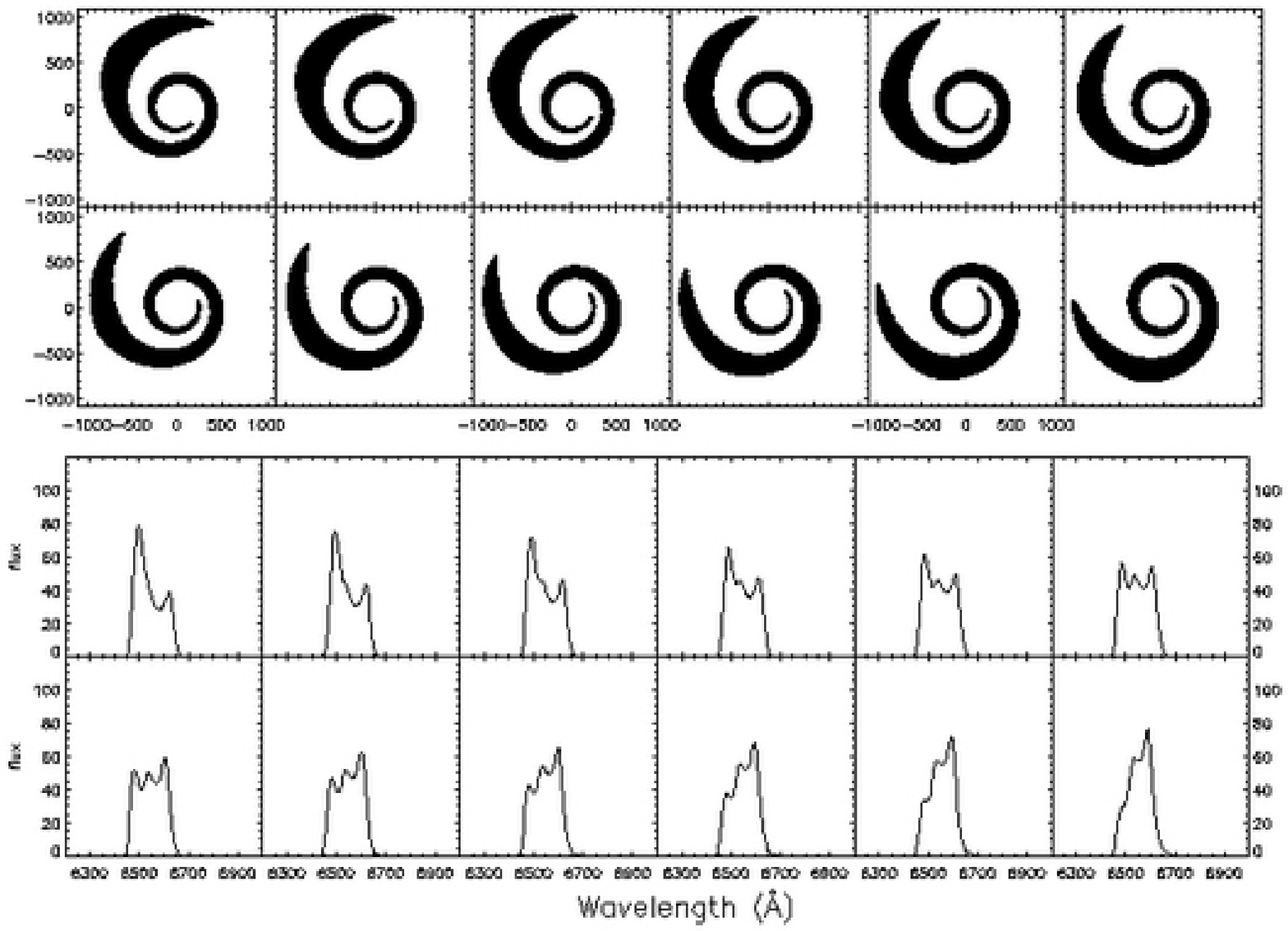}
  \caption{The images~(upper panel) of the illuminated area of the disc
and H$\alpha$ line profiles~(lower panel) computed by our code for twist
warped disc cases for $i=10^{\circ},\; n_1=2.6\pi,\;n_2=3.0,\; n_3=22^{\circ}$,
the disc zone is from $R_{\rm in}=250r_{\rm g}$ to $R_{\rm out}= 
1000 r_{\rm g}$ for a prograde disc. The longitude of the observer $\gamma_0$ 
with respect to the coordinate system of the disc from top left-hand side to 
bottom right-hand side takes steps of $10^{\circ}$ from $30^{\circ}$ to 
$140^{\circ}$. The variation of the line profiles with respect to different 
longitude corresponds to different observation time and can be compared with 
the future observations of source SDSS J232721.96+152437.3.
 \label{twist4}}
%\end{figure*}

%\begin{figure*}[htp]
\includegraphics[width=0.75\hsize]{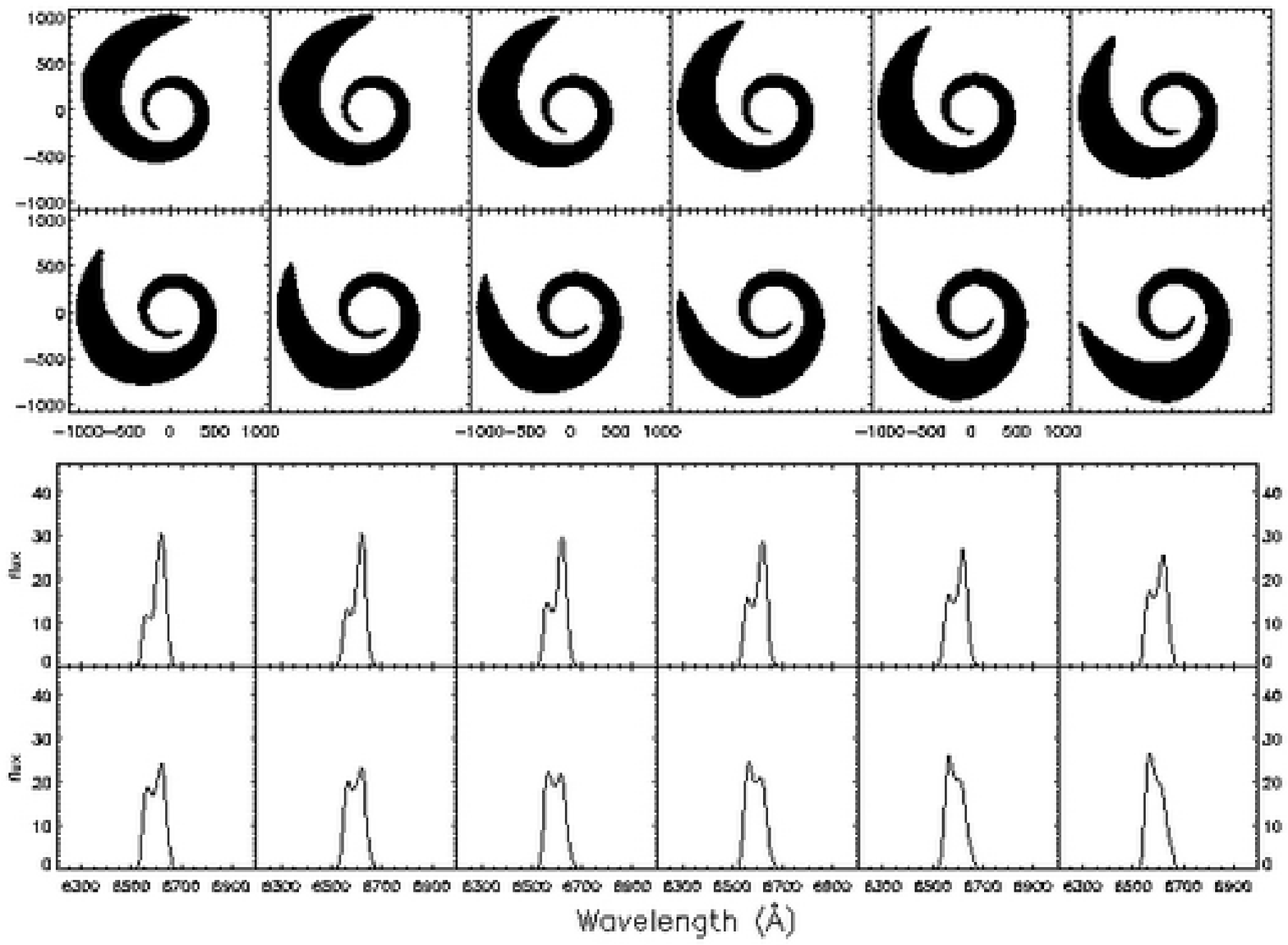}
  \caption{The images~(upper panel) of the illuminated area of the disc and 
H$\alpha$ line profiles~(lower panel) computed by our code for twist warped 
disc cases for $i=8^{\circ},\; n_1=2.1\pi,\;n_2=3.0,\; n_3=6^{\circ}$, the 
disc zone is from $R_{\rm in}=250r_{\rm g}$ to $R_{\rm out}= 1000 r_{\rm g}$ 
for a retrograde disc. The longitude of the observer $\gamma_0$ with respect 
to the coordinate system of the disc from top left-hand side to bottom 
right-hand side takes steps of $10^{\circ}$ from $50^{\circ}$ to $160^{\circ}$. 
The variation of the line profiles with respect to different longitude can be 
compared with successive observation of source SDSS J093653.84+533126.8.
 \label{twist5}}
\end{figure*}

The line profiles with respect to different longitude of the observer are 
corresponding to the observations in different times in reality. Thus the 
variation of the line profiles with respect to different longitude can be 
compared with successive observations. The images of the illuminated area 
and the H$\alpha$ line profiles for sources SDSS J232721.96+152437.3, 
SDSS J093653.84+533126.8 are shown in Figs~\ref{twist4} and \ref{twist5}. 
In Fig.~\ref{twist4} the parameters of the disc are: $i=10^{\circ},\; n_1 = 
2.6\pi,\; n_2=3.0,\; n_3=22^{\circ}$, $R_{\rm in}=250r_{\rm g}$, $R_{\rm out} 
=  1000 r_{\rm g}$ for a prograde disc. The longitude of the observer 
$\gamma_0$  from top left-hand side to bottom right-hand side takes steps 
of $10^{\circ}$ from $30^{\circ}$ to $140^{\circ}$. A retrograde disc with 
$i=8^{\circ},\; n_1=2.1\pi,\;n_2=3.0,\; n_3=6^{\circ}$, $R_{\rm in} = 
250r_{\rm g}$ and $R_{\rm out} = 1000 r_{\rm g}$, the longitude of the 
observer $\gamma_0$ from top left-hand side to bottom right-hand side takes 
steps of $10^{\circ}$ from $50^{\circ}$ to $160^{\circ}$ (shown in 
Fig.~\ref{twist5}). Long term monitoring of these sources should provide 
critical tests for the disc warping scenarios.

\section{Summary}

\label{sum} 
We compute the reprocessing emission-line profiles from a warped 
relativistic accretion disc around a central black hole by including all
relativistic effects.  A parametrized disc model is used to obtain an 
insight into the impact of disc warping on the double-peaked Balmer 
emission-line profiles. For simplicity, we assumed that the disc is 
illuminated by a point-like central source, which is a good approximation, 
the line emissivity is proportional to the continuum light intercepted by 
the accretion disc, and line emission is isotropic.   We find the following.
\begin{enumerate}
\item For twist-free or low phase amplitude~(described by $n_1$)  
warped disc, the asymmetrical and frequency-shifted single-peaked line 
profiles as expected are produced in most cases. The rarity of such 
sources suggests that a warping disc is usually twisted.  
\item The flux ratio of the blue peak to red one becoming less than unity 
can be predicted by a twisted warp disc with high phase amplitude. The 
phase amplitude has a significant influence on the line profiles.
\item A third peak and the variation in blue/red peak position which have
been found by observations \citep{vei91,str03} are produced in our model.
\item The influence from retrograde disc on line profiles has a inverse
effect compared with a prograde one.
\item The fraction of the radiation incident to the outer disc from the 
inner part can be enhanced by disc warping. The results are shown in 
Fig.~\ref{sang}.
\end{enumerate}

We showed that warped disc model is flexible enough to reproduce a variety of 
line profiles including triplet-peaked line profiles, or double peaked profiles 
with additional plateaus in the line wing as observed in the SDSS spectra of 
some AGN, while the models have the same number of free parameters as 
eccentric disc models. While we leave the detailed fit to a future paper, 
the essential characteristics in the line profiles is in good agreement with 
observed line profiles. Future monitoring of line profile variability in 
these triplet sources can provide a critical test of the warped disc model.

\section{acknowledgments}
We would like to thank the anonymous referee for his/her helpful suggestions 
and comments which improve and clarify our paper.

\label{lastpage}

\begin{thebibliography}{}

\bibitem[Armitage \& Natarajan (1999)]{arm99} Armitage P.~J.,
        Natarajan P., 1999, ApJ, 525, 909
\bibitem[Bardeen \& Petterson(1975)]{bap75} Bardeen J. M., Petterson J. A., 
      1975, \apj, 195, L65
\bibitem[Bardeen et al.(1972)Bardeen, Press \& Teukolsky]{bar72} Bardeen J. M., Press
    W. H., Teukolsky S. A., 1972, \apj, 178, 347
%\bibitem[Blackman(1999)]{bla99} Blackman E. G., 1999, \mnras, 306, L25
\bibitem[Bachev(1999)]{bac99} Bachev R., 1999, \aap, 348, 71
\bibitem[Begelman, Blandford \& Rees(1980)]{beg80} Begelman M. C., Blandford R. D., 
      Rees M. J., 1980, \nat, 287, 307
\bibitem[Bian et al.(2007)]{bian07} Bian W.-H., Chen Y.-M., Gu Q.-S., Wang J.-M.,
      2007, ApJ, 668, 721
\bibitem[\v{C}ade\v{z} \etal(2003)]{cad03} \v{C}ade\v{z} A., Brajnik M.,
     Gomboc A., Calvani M., Fanton C., 2003, \aap, 403, 29
\bibitem[\v{C}ade\v{z} et al.(1998)\v{C}ade\v{z}, Fanton \& Calvani]{cad98}
    \v{C}ade\v{z} A., Fanton C., Calvani M., 1998, New Astronomy, 3, 647
%\bibitem[Caproni et al.(2006)]{cap06} Caproni A., Livio M.,
%      Abraham Z., Mosquera Cuesta H. J., 2006, \apj, 653, 112
\bibitem[Caproni et al.(2007)]{cap07} Caproni A., Abraham Z.
       Livio M., Mosquera Cuesta H. J., 2007, \mnras, 379,135
\bibitem[Cao \& Wang(2006)]{cao06} Cao X., Wang T. G., 2006, \apj, 652, 112
\bibitem[Carter(1968)]{car68} Carter B., 1968, Phys. Rev., 174, 1559
\bibitem[Chakrabarti \& Wiita(1993)]{cha93} Chakrabarti S., Wiita P.~J.,
    1993, ApJ, 411, 602
\bibitem[Chakrabarti \& Wiita(1994)]{cha94} Chakrabarti S., Wiita P.~J.,
    1994, ApJ, 434, 518
\bibitem[Chen \& Halpern(1989)]{che89} Chen K., Halpern J.~P., 1989, ApJ, 344, 115
\bibitem[Chen et al.(1989)Chen, Halpern \& Filippenko]{chen89} Chen K.~, Halpern J.~P., 
    Filippenko A.~V., 1989, ApJ, 339, 742
\bibitem[Cunningham(1975)]{cun75} Cunningham C. T., 1975, \apj, 202, 788
\bibitem[Cunningham \& Bardeen(1973)]{cun73} Cunningham C. T., Bardeen
      J. M., 1973, \apj, 183, 237
\bibitem[Dong et al.(2008)]{2008MNRAS} Dong X., Wang T., Wang J., Yuan W.,
      Zhou H., Dai H., Zhang K., 2008, \mnras, 383, 581
\bibitem[Eracleous \& Halpern(1994)]{era94} Eracleous M.,
      Halpern J.~P., 1994, ApJS, 90, 1
\bibitem[Eracleous \& Halpern(2003)]{era03} Eracleous M., 
      Halpern J.~P., 2003, \apj, 599, 886
\bibitem[Eracleous \etal(1995)]{era95} Eracleous M.~, Livio M.~, Halpern J.~P.,
        Storchi-Bergmann T., 1995, ApJ, 438, 610
\bibitem[Eracleous \etal (1997)]{era97} Eracleous M., Halpern J.~P.,
      Gilbert A.~M., Newman J.A., Filippenko A.~V.,~1997, ApJ, 490, 21
\bibitem[Gezari et al.(2007)Gezari, Halpern \& Eracleous]{gez07} Gezari S., 
       Halpern J.~P., Eracleous M., 2007, ApJS, 169, 167
\bibitem[Gaskell(1996)]{gas96} Gaskell C. M., 1996, \apj, 464, 107
\bibitem[Hartnoll \& Blackman(2000)]{har00} Hartnoll S. A., Blackman E. G.,
     2000, \mnras, 317, 880
\bibitem[Herrnstein et al.(2005)]{her05} Herrnstein J. R., Moran J. M., Greenhill
     L. J., Trotter A., 2005, \apj, 629  719
\bibitem[Kumar \& Pringle(1985)]{kum85} Kumar S., Pringle J. E., 1985,
      \mnras, 213, 435.
\bibitem[Lai(1999)]{lai99} Lai, D. 1999, \apj, 524, 1030
\bibitem[Lai(2003)]{lai03} Lai, D. 2003, \apj, 591, L119
\bibitem[Larwood et al.(1996)]{lar96} Larwood J. D., Nelson R. P., Papaloizou
       J. C. B., Terquem C., 1996, \mnras, 282, 597
\bibitem[Lu \& Yu(2001)]{lu01} Lu Y., Yu Q., 2001, \apj, 561, 660
\bibitem[Maloney, Begelman \& Pringle(1996)]{mal96} Maloney P. R.,
       Begelman M. C., Pringle J. E., 1996, \apj, 472, 582
\bibitem[Maloney \& Begelman(1997)]{mal97} Maloney P. R., Begelman M. C., 1997,
       \apj, 491, L43
\bibitem[Maloney, Begelman \& Nowak(1998)]{mal98} Maloney P. R., Begelman M. C.,
       Nowak M. A., 1998, \apj, 504, 77
\bibitem[Miller \& Peterson(1990)]{mil90} Miller J.~S., 
         Peterson B.~M., 1990, ApJ, 361, 98
\bibitem[Nayakshin(2005)]{nay05} Nayakshin S.~, 2005, \mnras, 359, 545
\bibitem[Nelson \& Papaloizou(1999)]{nel99} Nelson R. P.,
      Papaloizou J. C. B., 1999, \mnras, 309, 929
\bibitem[Nelson \& Papaloizou(2000)]{nel00} Nelson R. P.,
      Papaloizou J. C. B., 2000, \mnras, 315, 570
\bibitem[Ogilvie(1999)]{ogi99} Ogilvie G. I., 1999, \mnras, 304, 557
%\bibitem[Ogilvie \& Dubus(2001)]{ogi01} Ogilvie G. I., Dubus G., 2001,
%      \mnras, 320, 485
\bibitem[Papaloizou \& Lin(1995)]{pal95} Papaloizou J. C. B., Lin D. N. C.,
       1995, \apj, 438, 841
\bibitem[Papaloizou \& Pringle(1983)]{pap83} Papaloizou J. C. B., Pringle J. E.,
       1983, \mnras, 202, 1181
%\bibitem[Papaloizou \& Terquem(1995)]{pap95} Papaloizou J. C. B., Terquem C., 1995,
%       \mnras, 274, 987
\bibitem[Pfeiffer \& Lai(2004)]{pfl04} Pfeiffer H. P., Lai D., 2004,
       \apj, 604, 766
\bibitem[Press et al.(1992)]{pre92} Press W. H., Teukolsky S. A., Vetterling
    W. T., Flannery B. P., 1992, Numerical Recipes. Cambridge University
    Press, Cambridge
\bibitem[Pringle(1996)]{pri96} Pringle J. E., 1996, \mnras, 281, 357
\bibitem[Pringle(1997)]{pri97} Pringle J. E., 1997, \mnras, 292, 136
\bibitem[Storchi-Bergmann \etal (1997)]{sto97} Storchi-Bergmann T.~,
       Eracleous M., Ruiz M.~T.~, Livio M.~, Wilson A.~S.~,
       FIlippenko A.~V.,~1997, ApJ, 489, 87
\bibitem[Strateva et al.(2003)]{str03} Strateva I. V. et al., 2003, AJ, 126, 1720
\bibitem[Terquem \& Bertout(1993)]{teb93} Terquem C., Bertout C., 1993, \aap, 274, 291
\bibitem[Terquem \& Bertout(1996)]{teb96} Terquem C., Bertout C., 1996, \aap, 279, 415
%\bibitem[Terquem \& Papaloizou(2000)]{tep00} Terquem C., Papaloizou J. C. B., 
%      2000, \aap, 360, 1031
\bibitem[Veilleux \& Zheng (1991)]{vei91} Veilleux S., Zheng W.,
       1991, ApJ, 377, 89
\bibitem[Viergutz(1993)]{vie93} Viergutz S. U., 1993, \aap, 272, 355
\bibitem[Wang et al.(2005)]{wang05} Wang T.-G., Dong X.-B., Zhang X.-G., 
       Zhou H.-Y., Wang J.-X., Lu Y.-J, 2005, \apj, 625, L35
%\bibitem[Wijers \& Pringle(1999)]{wij99} Wijers R. A. M. J., Pringle J. E., 1999,
%      \mnras, 308, 207
\bibitem[Wilkins(1972)]{wil72} Wilkins D. C., 1972, Phys. Rev. D5, 814
\bibitem[Wu \& Wang(2007)]{wu07} Wu S.-M., Wang T.-G., 2007, \mnras, 378, 841
\bibitem[Zhang, Dultzin-Hacyan \& Wang(2007a)]{zha07A} Zhang X. -G.,
      Dultzin-Hacyan D., Wang T.-G., 2007, \mnras, 376, 1335
\bibitem[Zhang, Dultzin-Hacyan \& Wang(2007b)]{zha07b} Zhang X. -G.,
      Dultzin-Hacyan D., Wang T.-G., 2007, \mnras, 377, 1215
\bibitem[Zheng, Sulentic \& Binette(1990)]{zhe90} Zheng W., Sulentic J. W., Binette L.,
      1990, \apj, 365, 115
\bibitem[Zhou et al.(2006)]{2006ApJS..166..128Z} Zhou H., Wang T., Yuan W., 
      Lu H., Dong X., Wang J., Lu Y., 2006, \apjs, 166, 128

\end{thebibliography}
\end{document}